\def\be{\begin{equation}}
\def\ee{\end{equation}}
\def\bea{\begin{eqnarray}}
\def\eea{\end{eqnarray}}
\def\bse{\begin{subequations}}
\def\ese{\end{subequations}}

\def\DT{D_{\text{T}}}

\documentclass[pre,twocolumn,amsmath,amssymb,eqsecnum,preprintnumbers,floatfix]{revtex4}
\usepackage{graphicx}
\usepackage{dcolumn}
\usepackage{bm}
\usepackage{bbm}
\usepackage{color}
\usepackage{enumitem}
\usepackage{soul}
\usepackage{bbold}
\begin{document}
\preprint{arXiv:24xx.yyyyy}
\preprint{PRE {\bf xxx}, yyyyyy (202x)}

\title{Fluctuation-Response Relation in Non-Equilibrium Systems and Active Matter}

\author{T.R. Kirkpatrick$^{1}$ and D. Belitz$^{2,3}$}

\affiliation{$^{1}$Institute for Physical Science and Technology, University of Maryland, College Park, MD 20742, USA\\
 $^{2}$Department of Physics and Institute for Fundamental Science, University of Oregon, Eugene, OR 97403, USA \\
 $^{3}$ Materials Science Institute, University of Oregon, Eugene, OR 97403, USA}

\date{\today}
\begin{abstract}
We use dynamic equations to derive a relation between correlation functions and response
or relaxation functions in many-body systems. The relation is very general and holds both
in equilibrium, when the usual fluctuation-dissipation theorem is valid, and to linear order
in an expansion about arbitrary non-equilibrium states, when it is not. We illustrate our
results by discussing fluids, both in equilibrium and non-equilibrium states, as well as
several active-matter systems. 
\end{abstract}

\maketitle

\section{Introduction}
\label{sec:I}

Important concepts in many-body systems include the correlations of spontaneous fluctuations of 
observables about a given state, and the response of observables to external perturbations. The 
former are described by time correlation functions; the latter, to linear order in the external fields, 
by (linear) response functions. The response functions also describe the energy dissipated by the
system \cite{Kadanoff_Martin_1963}. For systems in thermodynamic equilibrium both functions are 
translationally invariant with respect to both time and space, and there is a simple relation between 
them known as the fluctuation-dissipation theorem (FDT) \cite{Nyquist_1928, Callen_Welton_1951, Kubo_1966},
see Sec.~\ref{subsec:II.C} below. That is, the spontaneous fluctuations in equilibrium 
determine the linear response of the system to external fields, as well as its relaxation towards
equilibrium once the external fields have been turned off. The correlation functions are directly
measurable by light (or neutron) scattering. This can be difficult if low temperatures or/and
small scattering angles are required. The response or relaxation functions, by contrast, are measurable via
macroscopic perturbations that one has experimental control of irrespective of the temperature.
Apart from being of fundamental interest, the FDT thus provides an alternative way of probing
 correlations in equilibrium many-body systems. 

For systems that are not in thermal equilibrium, or for non-thermal systems that are not coupled
to a heat bath, the standard FDT does not hold. Examples include fluids in the presence of an externally 
imposed temperature gradient, fluids with turbulent flow, and active matter. Many of these display 
interesting correlations that can be very hard to probe directly. There is a large body of work on
non-equilibrium fluctuations (see, e.g., Refs.~\onlinecite{Sevick_et_al_2008, Seifert_2012, Gaspard_2022} and
references therein), on non-equilibrium linear response \cite{Cengio_Levis_Pagonabarraga_2019, Baiesi_Maes_2013}, 
and on relations between the two, with various applications \cite{Marconi_et_al_2008, 
Baiesi_Maes_Wynants_2009, Andrieux_Gaspard_2004, Maes_2020, Caprini_Puglisi_Sarracino_2021, 
Cengio_Levis_Pagonabarraga_2021, Gaspard_2022, Davis_Proesmans_Fodor_2024}. 
Generally, such relations are desirable since
the response of a system to external perturbations is usually relatively easy to control and measure. 
In this paper we point out a particularly simple relation between correlation functions and bilinear
products of linear response functions in non-equilibrium systems.  This fluctuation-response relation 
takes the form of a structural isomorphism rather than an identity. In addition to its simplicity, its 
significance is that it {\it always} holds to linear order in an expansion about an arbitrary state, 
irrespective of how badly the standard FDT is violated. It thus provides an alternative way of 
probing correlations that is valid very generally.




This paper is organized as follows. In Sec.~\ref{sec:II} we define the relevant correlation and 
response functions, and discuss the FDT. In Sec.~\ref{sec:III} we derive an isomorphism between
correlation functions and bilinear response functions that is generally valid and the main
result of the paper. In Sec.~\ref{sec:IV} we discuss various examples, and in Sec.~\ref{sec:V}
we conclude with a summary and a general discussion.

\section{Correlation functions and response functions}
\label{sec:II}

Consider observables $A_i(x)$ $(i=1,2, \ldots)$, where $x = ({\bm x},t)$ comprises spatial position ${\bm x}$
and time $t$. Let the average values be $\langle A_i\rangle$, and let $\delta A_i = A_i - \langle A_i\rangle$.
In a classical system, $\langle \ldots \rangle$ is a statistical mechanics average taken with an appropriate 
distribution function; in a quantum mechanical system, it denotes a quantum mechanical expectation value 
plus a statistical mechanics average. In an equilibrium state the averages are constants independent of 
space and time; in general they are space and time dependent. We define temporal and spatial Fourier transforms
\bse
\label{eqs:2.1}
\bea
\delta A_i({\bm x},\omega) &=& \int dt\,e^{i\omega t}\, \delta A_i(x)\ ,
\label{eq:2.1a}\\
\delta A_i({\bm k},t) &=& \int d{\bm x}\,e^{-i {\bm k}\cdot{\bm x}}\, \delta A_i(x)\ .
\label{eq:2.1b}
\eea
\ese
with ${\bm k}$ the wave vector and $\omega$ the frequency. We use $\int dt$ to
indicate an integral over all times, with analogous notation in temporal Fourier space, and 
$\int d{\bm x}$ a spatial integral over the system volume $V$.

In what follows we recall the definitions of correlation functions, which describe the correlations of the 
$\delta A_i$, and of response functions, which describe the response of $\langle A_i\rangle$ to external 
fields. Alternatively, the response functions describe the relaxation of the  $\langle A_i\rangle$ from initial 
values induced by the fields.

\subsection{Correlation functions}
\label{subsec:II.A}

In classical systems, the two-point correlation function is the van Hove function
\be
S_{ij}(x,x') =  \langle \delta A_i(x)\,\delta A_j(x')\rangle\ ,
\label{eq:2.2}
\ee
with analogous expressions in Fourier space. In quantum systems, one needs to distinguish between the
symmetrized, or anticommutator, correlation function
\be
C_{ij}(x,x') = \frac{1}{2} \langle \delta A_i(x)\,\delta A_j(x') + \delta A_j(x')\,\delta A_i(x)\rangle\ ,
\label{eq:2.3}
\ee
and the antisymmetrized, or commutator one \cite{double_prime_footnote}
\be
\chi_{ij}(x,x') = \frac{1}{2\hbar} \langle \delta A_i(x)\,\delta A_j(x') - \delta A_j(x')\,\delta A_i(x)\rangle\ .
\label{eq:2.4}
\ee

In the presence of space and time-translational invariance one has
\bse
\label{eqs:2.5}
\be
\varphi_{ij}(x,x') = \varphi_{ij}(x-x')\ ,
\label{eq:2.5a}
\ee
where $\varphi$ can stand for $S$, $C$, or $\chi$. In Fourier space this becomes
\be
\varphi_{ij}({\bm k},{\bm k}';\omega,\omega') = 2\pi \delta(\omega+\omega')\,V \delta_{{\bm k},-{\bm k}'}\,\varphi_{ij}({\bm k},\omega)\ ,
\label{eq:2.5b}
\ee
\ese
with obvious modifications if only space or time-translational invariance is present.

The relation between $S$, or $C$, and $\chi$ is complicated and known only for certain
special cases. In thermal equilibrium, when all of the correlation functions have structures
given by Eq.~(\ref{eq:2.5b}), the relation is \cite{Kadanoff_Martin_1963, Forster_1975}
\bse
\label{eqs:2.6}
\be
C_{ij}({\bm k},\omega) = \hbar\,\coth(\hbar\omega/2T)\,\chi''_{ij}({\bm k},\omega)\ ,
\label{eq:2.6a}
\ee
with $T$ the temperature. $\chi''$ is the spectral density of the commutator correlation
function as defined in Eq.~(\ref{eq:A.2d}). In the classical limit this simplifies to
\be
S_{ij}({\bm k},\omega) = (2T/\omega)\,\chi''_{ij}({\bm k},\omega)\ .
\label{eq:2.6b}
\ee
\ese
By integrating over all frequencies we obtain
\bse
\label{eqs:2.7}
\be
S_{ij}^0({\bm k}) = T \chi_{ij}^0({\bm k})\ .
\label{eq:2.7a}
\ee
Here $S^0$ is the equal-time van Hove function and
\be
\chi_{ij}^0({\bm x},{\bm x}') = \chi^+_{ij}({\bm x},{\bm x}';\omega=0) = \chi_{ij}({\bm x},{\bm x}';z=i\epsilon)
\label{eq:2.7b}
\ee
\ese
is the static commutator correlation function. Here and throughout the paper we denote by
\[
g^{\pm}(t) = g(t)\,\Theta(\pm t)\ ,
\]
with $\Theta$ the step function, the positive and negative-time parts of time-dependent functions 
$g(t)$, see Appendix~\ref{app:A}. The final expression in Eq.~(\ref{eq:2.7b}) is the Laplace transform
as defined in Eq.~(\ref{eq:A.2b}), with $\epsilon>0$ infinitesimal.
We emphasize that Eqs.~(\ref{eqs:2.6}, \ref{eq:2.7a}) by themselves represent relations between
two different types of correlation functions. Their usual interpretation as the FDT requires
the identification of $\chi$ with a response function. For non-equilibrium systems one needs
to carefully distinguish between commutator correlation functions and response functions,
and we will discuss this distinction in the following subsections.

\subsection{Response functions}
\label{subsec:II.B}

\subsubsection{The initial-value problem, and relaxation functions}
\label{subsubsec:II.B.1}

External fields or perturbations that couple to the observables will produce
a change $\delta\langle A_i\rangle$ of the average values: 
$a_i(x) \equiv \delta\langle A_i(x)\rangle = \langle A_i(x)\rangle  -  \langle A_i(x)\rangle_0$,
with $\langle\ldots\rangle_0$ the average in the absence of the perturbations.
Suppose the perturbations are switched off at $t=0$,
at which time they have created initial values $a_i^0({\bm x}) \equiv a_i({\bm x},t=0)$. Then for
$t>0$ the observables will relax toward their average values in the absence of the perturbations 
according to
\be
a_i^+(x) = \sum_j \int d{\bm x}'\,\phi_{ij}^+({\bm x},{\bm x}';t)\,a_j^0({\bm x}') \ .
\label{eq:2.8}
\ee
The $\phi_{ij}^+$ are relaxation functions \cite{Kubo_1957}.

\subsubsection{Response to an external field}
\label{subsubsec:II.B.2}

Let $h_i(x)$ be external fields that are conjugate to the variables in the sense that they couple 
linearly to the variables in a Hamiltonian or free-energy functional \cite{TDGL_footnote}. Then the 
change $a_i$ of $\langle A_i\rangle$ in response to the fields $h_j$, to linear order in the fields, is given by 
response functions $X_{ij}$ via \cite{Chaikin_Lubensky_1995, Forster_1975, DeGroot_Mazur_1984}
\be
a_i(x) = \int dx'\,\Theta(t-t')\,\sum_j\,X_{ij}(x,x')\,h_j(x')\ .
\label{eq:2.9}
\ee
Here $\int dx' = \int d{\bm x}' \int dt'$, and the step function ensures causality.

For many applications, $X_{ij}$ is time translationally invariant, at least in an approximate 
sense \cite{time_translation_invariance_footnote}. For the discussion of response functions, we will assume
that this is the case; generalizations will be the subject of a separate investigation. 
Equation~(\ref{eq:2.9}) then becomes
\bse
\label{eqs:2.10}
\be
a_i({\bm x},t)= \int dx'  \sum_j\,X_{ij}^+({\bm x},{\bm x}';t-t')\,h_j({\bm x}',t')
\label{eq:2.10a}
\ee
or, after a temporal Fourier transform,
\be
a_i({\bm x},\omega) = \int d{\bm x}' \sum_j X_{ij}^+({\bm x},{\bm x}';\omega)
\,h_j({\bm x}',\omega)\ .
\label{eq:2.10b}
\ee
\ese

\subsubsection{Relation between response and relaxation}
\label{subsubsec:II.B.3}

A relation between response and relaxation can be established as follows \cite{Kadanoff_Martin_1963, Forster_1975}. 
Consider fields that are switched on infinitely slowly in the distant past and switched off at time $t=0$:
\bse
\label{eqs:2.11}
\be
h_i({\bm x},t) = h_i({\bm x})\,e^{\epsilon t}\,\Theta(-t)\ ,
\label{eq:2.11a}
\ee
or
\be
h_i({\bm x},\omega) = h_i({\bm x})\,\frac{1}{i\omega + \epsilon}\ ,
\label{eq:2.11b}
\ee
\ese
with $\epsilon>0$ infinitesimal. For negative times,
Eq.~(\ref{eq:2.10a}) yields
\bse
\label{eqs:2.12}
\bea
a_i^{-}(x) &=& \sum_j \int dx' \, e^{-\epsilon t'}\,X^+_{ij}({\bm x},{\bm x}';t') \,h_j({\bm x}')\,e^{\epsilon t}
\nonumber\\
&=& \sum_j \int d{\bm x}'\,X_{ij}^0({\bm x},{\bm x}')\,h_j({\bm x}',t)\ ,
\label{eq:2.12a}
\eea
or
\be
a_i^-({\bm x},\omega) = \frac{1}{i\omega + \epsilon} \sum_j \int d{\bm x}'\,X_{ij}^0({\bm x},{\bm x}')\,h_j({\bm x}')\ ,
\label{eq:2.12b}
\ee
with
\be
X_{ij}^0({\bm x},{\bm x}') = X^+_{ij}({\bm x},{\bm x}';\omega=0) = X_{ij}({\bm x},{\bm x}';z=i\epsilon)
\label{eq:2.12.c}
\ee
\ese
the static response function. At time $t=0$ the fields have produced values of the observables given by
\be
a_i^0({\bm x}) = \sum_j \int d{\bm x}'\,X_{ij}^0({\bm x},{\bm x}')\,h_j({\bm x}')\ .
\label{eq:2.13}
\ee
For positive times, the $a_i$ will relax from these initial values, see Eq.~(\ref{eq:2.8}). From Eqs.~(\ref{eq:2.10b}), 
(\ref{eq:2.12b}), and (\ref{eq:2.13}) we obtain
\bse
\label{eqs:2.14}
\bea
a_i^+({\bm x},\omega) &=& \frac{1}{i\omega} \sum_j \int d{\bm x}' \left[X^+(\omega) (X^0
)^{-1} - 1\right]_{ij}({\bm x},{\bm x'})\,
\nonumber\\
&&\hskip 90pt \times a_j^0({\bm x}')\ .
\label{eq:2.14a}
\eea
where the term in square brackets is to be interpreted in a matrix sense. The relaxation functions
from Eq.~(\ref{eq:2.8}) are thus related to the response functions via
\be
\phi_{ij}^+({\bm x},{\bm x}';\omega) = \frac{1}{i\omega}\,\left[X^+(\omega) (X^0)^{-1} - 1\right]_{ij}({\bm x},{\bm x'})\ .
\label{eq:2.14b}
\ee
\ese

Note that Eq.~(\ref{eq:2.14a}) contains the same information as Eq.~(\ref{eq:2.10b}), but it requires
knowledge of the initial values of the observables only; not of the nature of the external fields that
have produced the initial values.

\subsection{The fluctuation-dissipation theorem and its validity}
\label{subsec:II.C}

In general, there is no obvious simple relation between the response functions $X$ and
the various correlation functions. However, in thermal equilibrium, when $C_{ij}$ and $\chi_{ij}$ 
have structures given by Eqs.~(\ref{eqs:2.5}), the commutator correlation functions determine 
the response functions via
\bse
\label{eqs:2.15}
\be
X^+_{ij}({\bm k},\omega) = \chi^+_{ij}({\bm k},\omega)\ .
\label{eq:2.15a}
\ee
Given this identity, it is natural to define response functions at negative times \cite{causality_footnote} via
\be
X^-_{ij}({\bm k},\omega) := \chi^-_{ij}({\bm k},\omega)\ .
\label{eq:2.15b}
\ee
The Laplace transforms in the negative half plane, and the spectral densities $\chi''$ and $X''$, 
for both $\chi$ and $X$ are then defined via Eqs.~(\ref{eqs:A.2}) and we have
\be
X_{ij}''({\bm k},\omega) = \chi_{ij}''({\bm k},\omega) = -\chi_{ji}''({\bm k},-\omega)\ .
\label{eq:2.15c}
\ee
and in particular
\be
X_{ij}^0({\bm k}) = \chi_{ij}^0({\bm k})\ .
\label{eq:2.15d}
\ee
\ese
The relations (\ref{eqs:2.6}, \ref{eq:2.7a}) then turn into the FDT. That is, they relate fluctuations,
as described by the correlation functions $S$ or $C$, to the response to external perturbations, and
the related dissipation, as described by $X = \chi$. For the purposes of this paper we just state the
FDT for classical systems:
\bse
\label{eqs:2.16}
\bea
S_{ij}({\bm k},\omega) = (2T/\omega)\,X''_{ij}({\bm k},\omega)\ .
\label{eq:2.16a}\\
S_{ij}^0({\bm k}) = T X_{ij}^0({\bm k})\ .
\label{eq:2.16b}
\eea
\ese

The above statements in general hold only for systems in thermal equilibrium. For non-equilibrium
systems Eq.~(\ref{eq:2.15a}) does in general not hold, which invalidates Eq.~(\ref{eqs:2.16}). Other aspects
of the FDT also break down; for instance, the $\coth$ factor in Eq.~(\ref{eq:2.6a}) relies both on a constant
temperature throughout the system and on the distribution function being the equilibrium one, but for the
examples we will consider the breakdown of Eq.~(\ref{eq:2.15a}) is the most crucial aspect of non-equilibrium.

In Secs.~\ref{sec:III} and \ref{sec:IV}  we will  show that nonetheless there still is a simple relation between 
correlation functions and response functions in non-equilibrium systems, and also in non-thermal systems
where no temperature concept exists.

\section{Dynamic equations}
\label{sec:III}

We now turn to the dynamic equations that govern the time evolution of observables. Consider an arbitrary, 
in general time dependent, state. Linearizing about this state \cite{linearizing_footnote} leads to an equation 
of motion of the form
\bse
\label{eqs:3.1}
\bea
\partial_t\, \delta A_i(x) &+& \sum_j  \int dx'\,\lambda^{\text r}_{ij}(x,x')\,\delta A_j(x') 
\nonumber\\
&=&
     \sum_j \int dx'\,b_{ij}(x,x')\,h_j(x') + f_i(x)\ .\qquad\ \ 
\label{eq:3.1a}
\eea
The matrix $\lambda_{ij}$ contains couplings between the variables, as well as relaxation rates, 
or memory functions, and 
\be
\lambda^{\text r}_{ij}(x,x') = \Theta(t-t')\,\lambda_{ij}(x,x')
\label{eq:3.1b}
\ee
\ese
is the retarded part of these functions. The $f_i$ in Eq.~(\ref{eq:3.1a}) are stochastic, or Langevin, forces whose 
average is zero. For a system in thermal equilibrium they describe thermal fluctuations. More generally, 
they describe noise that is inherent to the dynamics of the system, such as the random stirring in a stirred 
fluid, or imprecisions in the propagation of particles in active matter. We will assume them to be Gaussian
distributed with second moments
\bse
\label{eqs:3.2}
\be
\langle f_i(x)\,f_j(x')\rangle = \Delta_{ij}(x,x')\ ,
\label{eq:3.2a}
\ee
or, in temporal Fourier space,
\be
\langle f_i({\bm x},\omega)\,f_j({\bm x}',\omega')\rangle = \Delta_{ij}({\bm x},{\bm x}';\omega,\omega')\ .
\label{eq:3.2b}
\ee
\ese
The $b_{ij}$ describe the coupling of the dynamic variables to the external fields. They will be discussed 
in Sec.~\ref{subsubsec:III.B.2}. 

Note that the system of equations~(\ref{eq:3.1a}) is linear by construction; the average values
$\langle A_i\rangle$ that one expands about are in general described by nonlinear equations.

\subsection{Correlations}
\label{subsec:III.A}

In the absence of external fields we formally solve Eq.~(\ref{eq:3.1a}) for $\delta A$ and express
the correlation functions in terms of the correlations of the Langevin forces. For simplicity, we will 
consider classical systems unless otherwise noted; a generalization to anticommutator correlations
instead of van Hove correlations is straightforward. We define a temporal Fourier transform of 
$\lambda^{\text{r}}$ by
\be
\lambda_{ij}^{\text{r}}({\bm x},{\bm x}';\omega,\omega') = \int dt\,e^{i\omega t} \int dt'\,e^{-i\omega' t'}\,\lambda_{ij}^{\text{r}}(x,x')\ .
\label{eq:3.3}
\ee
In the absence of fields, Eq.~(\ref{eq:3.1a}) then becomes
\bse
\label{eqs:3.4}
\be
\sum_j \int d{\bm x}' \int d\omega'\,G_{ij}^{-1}({\bm x},{\bm x}';\omega,\omega')\,\delta A_j({\bm x}',\omega') = f_i({\bm x},\omega)\ .
\label{eq:3.4a}
\ee
Here
\bea
G_{ij}^{-1}({\bm x},{\bm x}';\omega,\omega') &=& -i\omega\, \delta_{ij}\,\delta({\bm x}-{\bm x}')\,\delta(\omega - \omega') 
\nonumber\\
&& + \lambda_{ij}^{\text{r}}({\bm x},{\bm x}';\omega,\omega')\ ,
\label{eq:3.4b}
\eea
is the inverse of a propagator $G$ that is defined by
\bea
\sum_l \int d{\bm x}'' && \hskip -20pt \int d\omega''\, G_{il}^{-1}({\bm x},{\bm x}'';\omega,\omega'') \,G_{lj}({\bm x}'',{\bm x}';\omega'',\omega')  
\nonumber\\
&=&  \delta_{ij}\, \delta({\bm x}-{\bm x}')\,\delta(\omega - \omega')\ .
\label{eq:3.4c}
\eea
\ese
The formal solution of Eq.~(\ref{eq:3.4a}) thus is
\be
\delta A_i({\bm x},\omega) = \sum_j \int d{\bm x}' \int d\omega'\, G_{ij}({\bm x},{\bm x}';\omega,\omega')\,f_j({\bm x}',\omega')\ .
\label{eq:3.5}
\ee
For the matrix of correlation functions this yields
\be
S_{ij}({\bm x},{\bm x}';\omega,\omega') = \left[G\,\Delta\,G^{\text{T}}\right]_{ij}({\bm x},{\bm x}';\omega,\omega')\ ,
\label{eq:3.6}
\ee
with the expression in brackets to be interpreted as a matrix product, and $G^{\text{T}}$ the transposed 
propagator \cite{propagator_footnote}. In quantum systems, one obtains the same structure for the anticommutator 
correlation function $C$.

Equation~(\ref{eq:3.6}) holds in complete generality. If we assume time translational invariance we have
\bse
\label{eqs:3.7}
\be
\langle f_i({\bm x},t)\,f_j({\bm x}',t')\rangle = \Delta_{ij}({\bm x},{\bm x}';t-t')
\label{eq:3.7a}
\ee
or
\be
\langle f_i({\bm x},\omega)\,f_j({\bm x}',\omega')\rangle = 2\pi\delta(\omega + \omega')\Delta_{ij}({\bm x},{\bm x}';\omega)\rangle\ ,
\label{eq:3.7b}
\ee
\ese
and
\bse
\label{eqs:3.8}
\be
G_{ij}({\bm x},{\bm x}';\omega,\omega') = \delta(\omega - \omega')\,G_{ij}({\bm x},{\bm x}';\omega)
\label{eq:3.8a}
\ee
with $G$ the inverse of
\be
G_{ij}^{-1}({\bm x},{\bm x}';\omega) = -i\omega\,\delta_{ij} \delta({\bm x}-{\bm x}') + \lambda^{+}_{ij}({\bm x},{\bm x}';\omega)\ .
\label{eq:3.8b}
\ee
\ese
Note that $\lambda^+(\omega) = \lambda(z=\omega+i\epsilon)$, see Eq.~(\ref{eq:A.2c}).
This yields for the correlation functions 
\be
S_{ij}({\bm x},{\bm x}';\omega) = \left[G(\omega)\,\Delta(\omega)\,G^{\dagger}(\omega)\right]_{ij}({\bm x},{\bm x}')\ ,
\label{eq:3.9}
\ee
with the adjoint $G^{\dagger}$ the complex conjugate of $G^{\text{T}}$.

If the system is translationally invariant in space as well, we have
\be
S_{ij}({\bm k},\omega) = \left[G({\bm k},\omega)\,\Delta({\bm k},\omega)\,G^{\dagger}({\bm k},\omega)\right]_{ij}\ .
\label{eq:3.10}
\ee

We note that in equilibrium there is no freedom for choosing the Langevin-force correlations $\Delta_{ij}$; they
are dictated by the FDT in the sense that Eq.~(\ref{eq:3.10}) must yields the correct equilibrium equal-time
correlation functions $S_{ij}^0$. More generally, the $\Delta_{ij}$ are part of what characterizes the 
nonequilibrium/nonthermal nature of the system.

\subsection{Linear Response}
\label{subsec:III.B}

To discuss the response, we average Eq.~(\ref{eq:3.1a}). Then the Langevin force drops out and
we obtain an equation for $\delta\langle A_i\rangle \equiv a_i$.
As in Sec.~\ref{subsec:II.B}, we first consider the initial-value problem.

\subsubsection{The initial-value problem}
\label{subsubsec:III.B.1}

Suppose initial values $a_i^0({\bm x})$ have been created by external perturbations. Then the
relaxation of the observables for positive times is governed by the averaged version of Eq.~(\ref{eq:3.1a}), viz.
\be
\partial_t\, a_i^+(x) + \sum_j \int dx'\,\lambda^{\text{r}}_{ij}(x,x')\,a_j^+(x') = 0\ .
\label{eq:3.11}
\ee
A temporal Fourier transform yields
\bea
-i\omega\,a_i^+({\bm x},\omega) \hskip -4pt &+& \hskip -4pt \sum_j \int \! d{\bm x}' \!\! \int \! d\omega'\,\lambda^{\text{r}}_{ij}({\bm x},{\bm x}';\omega,\omega')\,a_j^+({\bm x}',\omega')
\nonumber\\
     &=& a_i^0({\bm x})\ ,
\label{eq:3.12}
\eea
or
\be
a_i^+({\bm x},\omega) = \sum_j \int \! d{\bm x}' \!\! \int d\omega'\,G_{ij}({\bm x},{\bm x}';\omega,\omega')\,a_j^0({\bm x}')
\label{eq:3.13}
\ee
with the propagator $G$ as defined in Eqs.~(\ref{eqs:3.4}). The relaxation functions $\phi$ as defined in Eq.~(\ref{eq:2.8})
are thus given in terms of the propagator as
\be
\phi_{ij}^+({\bm x},{\bm x}';\omega) = \int d\omega'\,G_{ij}({\bm x},{\bm x}';\omega,\omega')\ .
\label{eq:3.14}
\ee

\subsubsection{Response to an external field}
\label{subsubsec:III.B.2}

Now explicitly consider a set of external fields. For this discussion we assume time translational invariance, either
exactly or approximately, as we did in Sec.~\ref{subsec:II.B}. Averaging Eq.~(\ref{eq:3.1a}) yields an equation for 
$a_i \equiv \delta \langle A_i\rangle$.
As in Sec.~\ref{subsubsec:II.B.3}
we consider fields that are switched on infinitely slowly in the distant past and switched off at time $t=0$, Eq.~(\ref{eqs:2.11}). 
Then $a_i^-$ is given by Eqs.~(\ref{eqs:2.12}). Adopting an obvious matrix notation, we write these equations as
\bse
\label{eqs:3.15}
\bea
\left\vert a^-(t)\right) &=& X^0 \left\vert h(t)\right)\ ,
\label{eq:3.15a}\\
\left\vert a^-(\omega)\right) &=& \frac{1}{i\omega + \epsilon}\,\left\vert a^0\right)\ ,
\label{eq:3.15b}
\eea
where
\be
\left\vert a^0\right) = X^0 \left\vert h(t=0)\right)\ .
\label{eq:3.15c}
\ee
\ese
For $t<0$ the time evolution of the observables thus follows the fields, and the equation of motion is
\be
\partial_t \left\vert a^-(t)\right) = \epsilon \left\vert a^-(t)\right)\ .
\label{eq:3.16}
\ee
For $t>0$ the relaxation rates $\lambda_{ij}$ cause the $a_i$ to relax to zero from the intial values created
by the fields. Using the past variables $a^-$ as the driving force, rather than the fields, we can therefore write 
the equation of motion for all times as
\be
\partial_t\,\vert a(t)) + \Theta(t)  \int_0^t dt'\,\lambda(t-t')\,\vert a(t')) = \epsilon\,  \vert a^-(t))\ .
\label{eq:3.17}
\ee
We now extend the $t'$-integral to all times by adding an integral over negative times to both
sides of the equation. After some simple manipulations, and again suppressing the infinite integration limits, this results in
\bea
\partial_t\,\vert a(t)) &+& \int dt'\,\lambda^+(t-t')\,\vert a(t')) 
\nonumber\\
&& \hskip 20pt = \int dt'\,\lambda^+(t-t')\,\vert a^-(t'))\ . \qquad
\label{eq:3.18}
\eea
A temporal Fourier transform yields
\bea
G^{-1}(\omega)\,\vert a(\omega)) &=& \lambda^+(\omega)\,\vert a^-(\omega))
\nonumber\\
&=& \left(G^{-1}(\omega) + i\omega\right) \frac{1}{i\omega + \epsilon}\, \vert a^0)\ .
\nonumber\\
\label{eq:3.19}
\eea
with $G^{-1}$ from Eq.~(\ref{eq:3.8b}). The solution of the initial-value problem thus is
\bse
\label{eqs:3.20}
\bea
\vert a(\omega)) &=& \frac{1}{i\omega}\,\left(1 + i\omega\,G(\omega)\right)\,\vert a^0)\ ,
\label{eq:3.20a}\\
\vert a^+(\omega)) &=& G(\omega)\,\vert a^0)\ .
\label{eq:3.20b}
\eea
\ese
Equation~(\ref{eq:3.20b}) is Eq.~(\ref{eq:3.13}) for the time-translationally invariant case. 

Alternatively, we can use Eq.~(\ref{eq:2.12a}) in the first line of (\ref{eq:3.19}) to write
\bse
\label{eqs:3.21}
\be
\vert a(\omega)) = G(\omega)\,b(\omega)\,\vert h(\omega))
\label{eq:3.21a}
\ee
with the elements of the matrix $b$ given by \cite{b_footnote}
\bea
b_{ij}({\bm x},{\bm x}';\omega) &=& \left[\lambda^+(\omega)\, X^0\right]_{ij}({\bm x},{\bm x}')
\nonumber\\
&=& \left[\left(G^{-1}(\omega) + i\omega\right) X^0\right]_{ij}({\bm x},{\bm x}')\ .\qquad\quad
\label{eq:3.21b}
\eea
\ese
By comparing with Eq.~(\ref{eq:2.9}) we find for the matrix of response functions
\bea
X^+_{ij}({\bm x},{\bm x}';\omega) &\equiv& X_{ij}({\bm x},{\bm x}';z=\omega + i\epsilon) 
\nonumber\\
&=& \left[G(\omega)\,b(\omega)\right]_{ij}({\bm x},{\bm x}') 
\nonumber\\
&=& \left[ \left(1 + i\omega\, G(\omega)\right) X^0\right]_{ij}({\bm x},{\bm x}')\ .\qquad
\label{eq:3.22}
\eea
Solving Eq.~(\ref{eq:3.22}) for $G$ we see that Eq.~(\ref{eq:3.20b}) is identical to 
Eq.~(\ref{eq:2.14a}), which was obtained by different but equivalent reasoning. 

If the system is translationally invariant in space we have
\bea
X^+_{ij}({\bm k},\omega) &=& \left[G({\bm k},\omega)\,b({\bm k},\omega)\right]_{ij}
\nonumber\\
                                       &=&  \left[ \left(1 + i\omega\, G({\bm k},\omega)\right) X^0({\bm k})\right]_{ij}\ .
\label{eq:3.23}
\eea

We stress again that the response problem and the initial-value problem are equivalent: Eqs.~(\ref{eqs:3.21})
describe the response of the observables to the external field, while Eqs.~(\ref{eqs:3.20}) describe the
relaxation of the observables from the initial condition created by the field. Note, however,
that the matrices $b$ and $X^0$ are in general not diagonal, and hence Eq.~(\ref{eq:3.21a}) may
contain couplings that are absent in Eq.~(\ref{eq:3.20a}). We will see an example of this in
Sec.~\ref{subsec:IV.C}.

\subsection{A general relation between response formulas and fluctuation formulas}
\label{subsec:III.C}

Now define a matrix of bilinear products of observables,
\be
{\cal A}_{ij}^{++}({\bm x},{\bm x}';\omega,\omega') = a_i^+({\bm x},\omega)\,a_j^+({\bm x}',\omega')\ .
\label{eq:3.24}
\ee
Equation~(\ref{eq:3.13}) yields
\bse
\label{eqs:3.25}
\be
{\cal A}_{ij}^{++}({\bm x},{\bm x}';\omega,\omega') = \left[G\,{\cal A}^{00}\,G^{\text{T}}\right]_{ij}({\bm x},{\bm x}';\omega,\omega')
\label{eq:3.25a}
\ee
where
\be
{\cal A}^{00}_{ij}({\bm x},{\bm x}') = a_i^0({\bm x})\,a_j^0({\bm x}')\ .
\label{eq:3.25b}
\ee
\ese

In the special case of time translational invariance we define
\bea
{\cal A}_{ij}^{++}({\bm x},{\bm x}';\omega) &=& {\cal A}_{ij}^{++}({\bm x},{\bm x}';\omega,-\omega) 
\nonumber\\
&=& \left[G(\omega)\,{\cal A}^{00}\,G^{\dagger}(\omega)\right]_{ij}({\bm x},{\bm x}')\ . \qquad
\label{eq:3.26}
\eea
By using Eq.~(\ref{eq:3.15c}) we can write, equivalently,
\bse
\label{eqs:3.27}
\be
{\cal A}_{ij}^{++}({\bm x},{\bm x}';\omega) = \left[G(\omega) X^0\, {\cal H}^{00}X^{0\,\text{T}} G^{\dagger}(\omega)\right]_{ij}({\bm x},{\bm x}')
\label{eq:3.27a}
\ee
where
\be
{\cal H}^{00}_{ij}({\bm x},{\bm x}') = h_i({\bm x})\,h_j({\bm x}')\ .
\label{eq:3.27b}
\ee
\ese

If in addition we have spatial translational invariance we define
\bse
\label{eqs:3.28}
\bea
{\cal A}_{ij}^{++}({\bm k},\omega) &=& {\cal A}_{ij}^{++}({\bm k},-{\bm k};\omega) 
\nonumber\\
&=&  \left[G({\bm k},\omega)\,{\cal A}^{00}({\bm k})\,G^{\dagger}({\bm k},\omega)\right]_{ij}\ . \qquad
\label{eq:3.28a}
\eea
with
\be
{\cal A}_{ij}^{00}({\bm k}) = {\cal A}_{ij}^{00}({\bm k},-{\bm k})\ .
\label{eq:3.28b}
\ee
\ese

Now compare the relaxation formulas, Eqs.~(\ref{eqs:3.25}), or the response formula,
Eqs.~(\ref{eqs:3.27}), with the corresponding fluctuation formula, Eq.~(\ref{eq:3.6}),
which we write down again for clarity:
 \be
S_{ij}({\bm x},{\bm x}';\omega,\omega') = \left[G\,\Delta\,G^{\text{T}}\right]_{ij}({\bm x},{\bm x}';\omega,\omega')\ ,
\tag{3.6}
\ee
We see that the response, or relaxation, formulas have the same structure as the fluctuation formula; 
the only difference is the matrix sandwiched between the propagators.
This isomorphism between the fluctuation formula and the relaxation/response formula is our main
result. It is very general and holds for linear deviations from arbitrary states. In particular, 
it holds for deviations from an equilibrium state, when the usual FDT holds, as we 
will see below. However, regardless of whether or not the system is in equilibrium, any 
interesting correlations are encoded in the propagator $G$, and $S_{ij}$ and ${\cal A}_{ij}^{++}$ 
just represent different linear combinations of the propagator elements. The correlations can 
thus be probed by response, or relaxation, experiments even in absence of the usual FDT. 
Physically, this reflects the fact that, to linear order, the spontaneous fluctuations, the relaxation 
from an initial condition, and the response to external fields contain the same information. This is a 
manifestation of Onsager's regression hypothesis \cite{Onsager_1931b}.
In the following section we will discuss various examples.

\section{Examples}
\label{sec:IV}

In this section we discuss various examples that illustrate the fluctuation-response relation
derived in Sec.~\ref{sec:III}. We first demonstrate that the isomorphism does indeed hold
in equilibrium, as it must, and is equivalent to the FDT, by considering shear diffusion 
in a fluid in equilibrium. In more complicated systems, including fluids in non-equilibrium steady 
states (NESS) \cite{Kirkpatrick_Cohen_Dorfman_1982a, Kirkpatrick_Cohen_Dorfman_1982c}, 
stirred fluids \cite{Forster_Nelson_Stephen_1977}, or active matter \cite{Marchetti_et_al_2013, active_matter_footnote},
the general fluctuation-response isomorphism still holds, but the FDT breaks down due
to one or both of the following mechanisms:
\medskip\par\noindent
{\bf Mechanism I:} The stochastic-force correlations are different from those required by the FDT.
This obviously invalidates the FDT, while the isomorphism between the response or relaxation
formulas and the fluctuation formula demonstrated in Sec.~\ref{subsec:IV.C} remains intact. It
is interesting to distinguish between to subcases:
\begin{description}
\item[\hskip 5pt (a)] The wave-number dependence of the random-force correlation $\Delta$ is the same
as in equilibrium. In this case the FDT is violated only by a constant factor, and the correlations
have the same structure as in equilibrium. Examples are active Brownian motion, see Sec.~\ref{subsubsec:IV.B.1}, 
and some models of randomly stirred fluids discussed in Sec.~\ref{subsubsec:IV.B.2} below.
\item[\hskip 5pt (b)] The wave-number dependence of $\Delta$ is different from the equilibrium case. This
can lead to long-ranged correlations that are not present in equilibrium. Examples are certain
models for randomly stirred fluids, Sec.~\ref{subsubsec:IV.B.2}, and models of cell motion in biological tissues,
see Sec.~\ref{subsubsec:IV.B.3}.
\end{description}
\medskip\par\noindent
{\bf Mechanism II:} The inverse propagator $G^{-1}$ in Eqs.~(\ref{eqs:3.2}) contains terms that are not present 
in ordinary linearized hydrodynamics. This will in general lead to a breakdown of the FDT as well as
to long-ranged correlations that are not related to the properties of the random forces. An example is
a fluid in a constant temperature gradient, see the discussion in Sec.~\ref{subsec:IV.C}.

\subsection{Fluid in equilibrium}
\label{subsec:IV.A}

Consider shear diffusion in a classical fluid (or a quantum fluid in the hydrodynamic regime) 
\cite{Landau_Lifshitz_VI_1966, Chaikin_Lubensky_1995, Forster_1975}. 
In this case there is only one relevant observable, viz., one component of the transverse fluid velocity, which 
decouples from the other hydrodynamic variables and which we denote by $u_{\perp}$. The system is
translationally invariant with respect to both time and space. The inverse propagator is given by
\bse
\label{eqs:4.1}
\be
G^{-1}({\bm k},\omega) = -i\omega + \nu k^2\ ,
\label{eq:4.1a}
\ee
with $\nu$ the kinematic viscosity. From Eq.~(\ref{eq:3.8b}) we see that $\lambda^+$ is
frequency independent and given by
\be
\lambda^+({\bm k}) = \nu k^2\ .
\label{eq:4.1b}
\ee
\ese
For a fluid in equilibrium the random-force correlation is \cite{fluctuating_hydrodynamics_footnote, equal_time_correlation_footnote}
\be
\Delta({\bm k},\omega) = \frac{2T}{\rho}\,\nu {\bm k}^2 \ ,
\label{eq:4.2}
\ee
with $\rho$ the mass density. For the van Hove function, Eq.~(\ref{eq:3.10}) yields
\be
S({\bm k},\omega) = \frac{2T}{\rho}\,\frac{\nu k^2}{\omega^2 + (\nu k^2)^2}\ .
\label{eq:4.3}
\ee

The static susceptibility is  \cite{Forster_1975, static_susceptibility_footnote}
\bse
\label{eqs:4.4}
\be
\chi^0 = \int (d\omega/\pi)\,\chi''({\bm k},\omega) = 1/\rho\ .
\label{eq:4.4a}
\ee
Since this is an equilibrium system, Eq.~(\ref{eq:2.15a}) holds. We have
\be
X^0 = \chi^0 = 1/\rho\ ,
\label{eq:4.4b}
\ee
\ese
and from Eqs.~(\ref{eq:3.22}), (\ref{eq:4.1a}) and (\ref{eq:4.4b}) we find 
\bse
\label{eqs:4.5}
\be
X^+({\bm k},\omega) = \chi^+({\bm k},\omega) =  \frac{1}{\rho}\,\frac{\nu k^2}{-i\omega + \nu k^2}\ .
\label{eq:4.5a}
\ee
As a commutator correlation function, $\chi$ is an odd function of the frequency, so
\be
\chi^-({\bm k},\omega) =  \frac{1}{\rho}\,\frac{\nu k^2}{i\omega + \nu k^2}\ ,
\label{eq:4.5b}
\ee
and the spectral densities are (see Eq.~(\ref{eq:A.2d}), and also Eqs.~(\ref{eq:2.15b}, \ref{eq:2.15c}))
\be
X''({\bm k},\omega) = \chi''({\bm k},\omega) = \frac{1}{\rho}\,\frac{\omega\,\nu k^2}{\omega^2 + (\nu k^2)^2}\ .
\label{eq:4.5c}
\ee
\ese
Comparing Eqs.~(\ref{eq:4.3}) and (\ref{eq:4.5c}) we see that the FDT, Eqs.~(\ref{eqs:2.16}), holds.

Now consider the bilinear response formula. From, Eq.~(\ref{eq:3.25b}) we have 
\bse
\label{eqs:4.6}
\bea
{\cal A}^{00}({\bm k}) &=& \vert u_{\perp}^0({\bm k})\vert^2
\label{eq:4.6a}\\
&=& \frac{1}{\rho^2}\,\vert h_{\perp}({\bm k})\vert^2
\label{eq:4.6b}
\eea
with $h_{\perp}$ the field conjugate to $u_{\perp}$. Equations~(\ref{eqs:3.28}) yield
\be
{\cal A}^{++}({\bm k},\omega) = \frac{1}{\omega^2 + (\nu k^2)^2}\,{\cal A}^{00}({\bm k})\ ,
\label{eq:4.6c}
\ee
\ese
which expresses the bilinear response in terms of the initial condition, or, alternatively, the
external field. Comparing ${\cal A}^{++}$ with the van Hove function, Eq.~(\ref{eq:4.3}), 
we explicitly see the isomorphism discussed in Sec.~\ref{subsec:III.C}.

In order to make contact with Ref.~\onlinecite{Kirkpatrick_Belitz_2024} we also note that if 
we define a closely related observable,
\bse
\label{eqs:4.7}
\bea
{\cal B}^{+}({\bm k},\omega) &=& a({\bm k},\omega)\,a^+(-{\bm k},-\omega)
\nonumber\\
 &&\hskip -40pt = \left[(1 + i\omega G({\bm k},\omega)) X^0\, {\cal H}({\bm k},\omega) (-i\omega) X^{0} G({\bm k},-\omega)\right]\ ,
\nonumber\\                                    
\label{eq:4.7a}
\eea
with
\be
{\cal H}({\bm k},\omega) = \vert h({\bm k},\omega)\vert^2\ ,
\label{eq:4.7b}
\ee
\ese
we find                                            
\be
{\cal B}^{+}({\bm k},\omega) = \frac{-i}{\rho}\,\chi''({\bm k},\omega)\,{\cal H}({\bm k},\omega)\ .
\label{eq:4.8}
\ee
In this form, the relation between the fluctuations, as represented by $\chi'' = (\omega/2T) S$, and the
bilinear response, as represented by ${\cal B}^+$, was formulated in
Ref.~\onlinecite{Kirkpatrick_Belitz_2024} (the quantity $u_{\perp}^+$ was denoted by
${\tilde u}_{\perp}$ in that reference). 

The above relations between the bilinear response functions ${\cal A}^{++}$, or ${\cal B}^+$, and
the correlation functions $S$, or $\chi''$, are equivalent to the usual FDT. As derived from the Langevin
equation (\ref{eq:3.1a}), they are actually more plausible, and easier to understand, than the FDT,
given that the fluctuating force and the external field are both linear inhomogeneities in Eq.~(\ref{eq:3.1a}),
and the correlation function is quadratic in the fluctuating force.

\subsection{Examples of Mechanism I}
\label{subsec:IV.B}

We now discuss various examples where the FDT breaks down due to the properties of the
fluctuating-force correlations, while the isomorphism explained in Sec.~\ref{subsec:IV.C} (see
Eqs.~(\ref{eqs:3.25}) and (\ref{eq:3.6})) remains valid.

\subsubsection{Active Brownian motion}
\label{subsubsec:IV.B.1}

Ordinary Brownian motion is the random walk experienced by a particle of mass $m$ suspended
in a fluid in equilibrium at temperature $T$ \cite{Forster_1975, Ma_1976}. The effect of the fluid on 
the particle is described by two forces: a frictional force that is proportional to the particle's velocity, 
and a stochastic or Langevin force. We also include an external force ${\bm h}$ that is conjugate 
to the velocity. The equation of motion for the velocity then is
\bse
\label{eqs:4.9}
\be
\frac{d}{dt}\,{\bm v}(t) + \gamma\,{\bm v}(t) = \frac{1}{m}\,\gamma\, {\bm h}(t)+ {\bm f}(t)\ .
\label{eq:4.9a}
\ee
Here $\gamma > 0$ is the friction coefficient and ${\bm f}$ is the Langevin force with correlations
\be
\langle f_i(t)\,f_j(t') \rangle = \delta_{ij}\,\delta(t-t')\,\Delta\ .
\label{eq:4.9b}
\ee
\ese
Equations~(\ref{eqs:4.9}) are a simplified (linearized, and considered in zero spatial dimensions) version
of Model A in Ref.~\onlinecite{Hohenberg_Halperin_1977}. The form of the external force term is consistent
with Eqs.~(\ref{eqs:3.21}); the static response function is 
\be
X_{ij}^0 = \delta_{ij}/m\ ,
\label{eq:4.10}
\ee
see also Eq.~(\ref{eq:4.4b}). The
conjugate force ${\bm h}$ is related to the physical force ${\bm F}$ on the particle by 
${\bm h} = {\bm F}/\gamma$ \cite{TDGL_footnote}.
The average kinetic energy of the particle, $E = m \langle {\bm v}(t)\cdot{\bm v}(t)\rangle/2$, is proportional
to the equal-time velocity autocorrelation function $S^0$. The equipartition theorem yields
\be
S_{ij}^0 = \delta_{ij}\,T/m
\label{eq:4.11}
\ee
and the FDT, Eq.~(\ref{eq:2.16b}), holds.
The inverse propagator is $G_{ij}^{-1}(\omega) = \delta_{ij}\,G^{-1}(\omega)$ with
\be
G^{-1}(\omega) = -i\omega + \gamma\ ,
\label{eq:4.12}
\ee
and the dynamic van Hove function $S_{ij}(\omega) = \delta_{ij}\,S(\omega)$ is
\bse
\label{eqs:4.13}
\be
S(\omega) = \frac{\Delta}{\omega^2 + \gamma^2}\ .
\label{eq:4.13a}
\ee
In order for Eqs.~(\ref{eq:4.11}) and (\ref{eq:4.13a}) to be consistent, the random-force
correlation in equilibrium must be \cite{equal_time_correlation_footnote}
\be
\Delta = \gamma\,T/2m\ .
\label{eq:4.13b}
\ee
\ese
The particle's mean-square displacement can be derived from the equation of motion for the position,
\be
\frac{d}{dt}\,{\bm r}(t) = {\bm v}(t)\ .
\label{eq:4.14}
\ee
Differentiating again with respect to $t$, inserting Eq.~(\ref{eq:4.9a}), multiplying by ${\bm r}$,
averaging, and integrating yields Langevin's result \cite{Langevin_1908}
\be
\langle {\bm r}^2(t)\rangle = \frac{6T}{m\gamma} \left[t - \frac{1}{\gamma}\left(1 - e^{-\gamma t}\right)\right]\ .
\label{eq:4.15}
\ee

To obtain the dynamic response function we average Eq.~(\ref{eq:4.9a}) and find 
$X^+_{ij}(\omega) = \delta_{ij}\,X^+(\omega)$ with
\bse
\label{eqs:4.16}
\bea
X^+(\omega) &=& \chi^+(\omega) = \frac{1}{m}\,\frac{\gamma}{-i \omega + \gamma}\ ,
\label{eq:4.16a}\\
X''(\omega) &=& \chi''(\omega) = \frac{1}{m}\,\frac{\gamma\omega}{\omega^2 + \gamma^2}\ ,
\label{eq:4.16b}
\eea
\ese
where the reasoning is the same as in Sec.~\ref{subsec:IV.A}.
Comparing Eqs.~(\ref{eqs:4.13}) and (\ref{eq:4.16b}) we see that the FDT in the dynamic form
(\ref{eq:2.16a}) also holds, as it must. 

`Active Brownian motion' refers to a Brownian particle equipped with an energy source that is independent
of the heat bath provided by the fluid; for a review, see Ref.~\onlinecite{Romanczuk_et_al_2012}. 
One way to model such a system is to make the friction coefficient velocity dependent such that it is
negative for small $v$ and changes sign at some $v_0>0$:
\be
\gamma \to -\gamma_1 + \gamma_2 v^2\ ,
\label{eq:4.17}
\ee
with $\gamma_1, \gamma_2 > 0$. The fluctuating-force correlations are still given by
Eq.~(\ref{eq:4.9b}), but $\Delta$ is not constrained by the equipartition theorem. 
This makes the system an example for Mechanism I\,(a). 
Equation (\ref{eq:4.17}) leads to a nonzero value of the velocity in the absence
of an external field, making the particle self-propelled. Let this spontaneous velocity point in the
$z$-direction,
\be
\langle{\bm v}\rangle_{h=0} = v_0\,\hat{\bm z}\ ,
\label{eq:4.18}
\ee
where $v_0 = \sqrt{\gamma_1/\gamma_2}$. Linearizing about this NESS we obtain an equation
of motion for $u_z = v_z - v_0$ \cite{uperp_footnote}
\be
\frac{d}{dt}\,u_z(t) + 2\gamma_1\,u_z = \frac{1}{m}\,2\gamma_1\,h_z(t) + f_z(t)\ .
\label{eq:4.19}
\ee
The problem now maps onto the equilibrium case with $\gamma \to 2\gamma_1$. However,
the FDT does not hold due to Mechanism I\,(a), since $\Delta$ is unconstrained and the
equal-time correlation function is not given by Eq.~(\ref{eq:4.11}), whereas the response
function, which does not depend on $\Delta$, is unchanged. The bilinear 
relaxation function is 
\be
{\cal A}^{++}(\omega) = \frac{(u_z^0)^2}{\omega^2 + 4\gamma_1^2}
\label{eq:4.20}
\ee
and the isomorphism between $S$ and ${\cal A}^{++}$ holds. 

\subsubsection{Randomly stirred fluid}
\label{subsubsec:IV.B.2}

Various models of a randomly stirred fluid were discussed in Ref.~\onlinecite{Forster_Nelson_Stephen_1977}
with a focus on turbulence and low dimensions. In their simplest form, a linearization of these models 
yields a forced diffusion equation for the fluid velocity as in Sec.~\ref{subsec:IV.A}, but with a 
random-force correlation that does not obey the FDT. Accordingly, the inverse propagator is given by 
Eq.~(\ref{eq:4.1a}). The random-force correlation is a frequency-independent quantity $\Delta({\bm k})$ 
that is {\em not} given by Eq.~(\ref{eq:4.2}), but either proportional to $k^2$ with an independent coefficient, 
or a constant that is nonzero in a certain wave-number range.  The former case is an example of
Mechanism I\,(a); the latter, of Mechanism I\,(b). The van Hove function is
\be
S({\bm k},\omega) = \frac{\Delta({\bm k})}{\omega^2 + (\nu k^2)^2}
\label{eq:4.21}
\ee
and the bilinear relaxation function ${\cal A}^{++}$ is given by Eq.~(\ref{eq:4.6c}). We see that the
isomorphism between $S$ and ${\cal A}^{++}$ still holds, but the FDT does not. 

For the static or equal-time correlation function, Eq.~(\ref{eq:4.21}) yields
\be
S^0({\bm k}) = \int \frac{d\omega}{2\pi}\,S({\bm k},\omega) = \frac{\Delta({\bm k})}{2\nu k^2}\ .
\label{eq:4.22}
\ee
In models where $\Delta(k=0)\neq 0$, Mechanism I\,(b) thus leads to long-range static correlations,
whereas the static response function is the same as in equilibrium, $X^0 = 1/\rho$.
This is in contrast to the example discussed in Sec.~\ref{subsec:IV.C} below, where the fluctuating-force
correlations are the same as in equilibrium, the long-range correlations
are encoded in the propagator via a coupling between two diffusive fluctuations,
and they manifest themselves both in the equal-time correlation functions, Eqs.~(\ref{eqs:4.32}),
and the static response functions, Eq.~(\ref{eq:4.33}).

\subsubsection{Cell motion in biological tissues}
\label{subsubsec:IV.B.3}

Another example of an active-matter system \cite{active_matter_footnote} is the model for cell motion 
in biological tissues driven by cell division and apoptosis studied in Ref.~\onlinecite{Li_et_al_2022}. 
The two relevant hydrodynamic variables are the density $\rho$ of the active matter and one component
of the transverse fluid velocity, $u_{\perp}$. The linearized model dynamic equations are
\bea
\partial_t\,\rho - D{\bm\nabla}^2 \rho &=& f_{\rho}\ ,
\nonumber\\
\partial_t\,u_{\perp} - \nu{\bm\nabla}^2 u_{\perp} &=& f_{\perp}\ ,
\label{eq:4.23}
\eea
with $D$ a diffusion coefficient and $\nu$ the kinematic viscosity of the fluid.
Accordingly, the inverse propagator is
\be
G^{-1}({\bm k},\omega) = \begin{pmatrix} 
                                                  -i\omega + D k^2  & 0                               \\
                                                                           0  &  -i\omega + \nu k^2             
                               \end{pmatrix}\ .
\label{eq:4.24}                               
\ee                               
The fluctuating-force correlations in this model are constants $\Delta_{\rho}$ and $\Delta_{\perp}$,
respectively, which makes this model an example of Mechanism I\,(b) and maps the linearized 
problem onto the randomly stirred fluid, Sec.~\ref{subsubsec:IV.B.2}, with a ${\bm k}$-independent
$\Delta$. The FDT breaks down, while $S$ and ${\cal A}^{++}$ are still
isomorphic.

We also note that taking into account the nonlinearities in the equations leads to a very strong
long-time tail: $D$ and $\nu$ become frequency dependent and diverge for $\omega\to 0$ as
$1/\omega^{1/5}$ in spatial dimensions $d=3$, and $1/\omega^{1/2}$ in $d=2$. In time space,
this corresponds to an autocorrelation function that decays for long times as $1/t^{4/5}$ in $d=3$ and
$1/t^{1/2}$ in $d=2$, respectively \cite{Li_et_al_2022}. This temporal long-ranged correlation, just
as the spatial one in Eq.~(\ref{eq:4.22}), is a result of Mechanism I\,(b), albeit one that manifests
itself only at the level of the nonlinear theory. In $d=2$ these strong long-time tails have been
observed experimentally, see Ref.~\onlinecite{Li_et_al_2022} and references therein.

\subsubsection{Flocking}
\label{subsubsec:IV.B.4}

A rather complex example of active matter is the Toner-Tu flocking theory \cite{Toner_Tu_1995, Toner_Tu_1998},
which is an effective continuum theory for the model proposed by Vicsek et al.  \cite{Vicsek_et_al_1995}; for 
reviews, see Refs.~\onlinecite{Toner_Tu_Ramaswamy_2005, Marchetti_et_al_2013}. The hydrodynamic variables
are the coarse-grained density $\rho({\bm x},t)$ of the flock, with average value $\rho_0$, and a velocity field 
${\bm v}({\bm x},t)$ that is the coarse-grained velocity of the members of the flock. In the isotropic, or disordered,
phase, where the mean velocity is zero, the linearized dynamic equations
are 
\bea
\partial_t \rho + \rho_0 {\bm\nabla}\cdot{\bm v} &=& 0\ ,
\nonumber\\
\partial_t v_i + \gamma v_i + \sigma_1 \partial_i \rho - D_{\text B} \partial_i \partial_j v_j - D_{\text{T}}{\bm\nabla}^2 v_i &=& f_i\ .
\nonumber\\
\label{eq:4.25}
\eea
Here $D_{\text{B}}$ and $D_{\text{T}}$ are diffusion coefficients that are analogous to linear combinations
of the shear and bulk viscosities in an equilibrium fluid, and the coefficient $\sigma_1$ arises from
an expansion of the pressure in powers of the density fluctuations. $\gamma > 0$ is a friction coefficient,
and the $f_i$ are random forces with correlations given by Eq.~(\ref{eq:4.9b}).

Consider for simplicity the transverse components of the velocity. The inverse propagator is
\be
G^{-1}({\bm k},\omega) = -i\omega + \gamma + \DT k^2\ .
\label{eq:4.26}
\ee
and the corresponding van Hove function is
\be
S({\bm k},\omega) = \frac{\Delta}{\omega^2 + (\gamma + \DT k^2)^2}\ .
\label{eq:4.27}
\ee
Note that the static correlation function is finite at ${\bm k}=0$, in contrast to the case
of a stirred fluid. This is a consequence of the constant damping term \cite{dry_wet_footnote}.
The flock is not a thermal system (there is no heat bath, and no temperature concept), so
the question of the validity of the FDT is not well posed. However, the discussion in 
Sec.~\ref{subsec:III.C} still applies and the relaxation formula is
\be
{\cal A}^{++}({\bm k},\omega) = \frac{1}{\omega^2 + (\gamma + \DT k^2)^2}\,{\cal A}^{00}({\bm k})
\label{eq:4.28}
\ee
with ${\cal A}^{00}$ given by Eq.~(\ref{eq:4.6a}). This demonstrates the isomorphism between
$S$ and ${\cal A}^{++}$ in the isotropic phase.

The self-propulsion capability of the members of the flock is modeled as in the active Brownian 
motion model, via Eq.~(\ref{eq:4.17}). The resulting phase with a nonzero average velocity is quite 
complex, and we will not discuss it here. The propagators and correlation functions for the linearized 
theory have been given in Refs.~\onlinecite{Toner_Tu_1998, Toner_Tu_Ramaswamy_2005},
and \onlinecite{Marchetti_et_al_2013}. Given the propagators, ${\cal A}^{++}$ is again given by 
Eqs.~(\ref{eqs:3.28}).

The spontaneously broken rotational symmetry leads to an equal-time velocity autocorrelation function 
that diverges in the long-wavelength limit \cite{Toner_Tu_1995}, in contrast to the isotropic phase. 
Nonetheless, the model yields true long-range order in spatial dimension $d=2$. This is a result
of the nonlinear terms in the theory, and consistent with numerical simulations \cite{Vicsek_et_al_1995}.

\subsection{An Example of Mechanism II: Fluid in a NESS}
\label{subsec:IV.C}

A simple example of a non-equilibrium system where Mechanism II leads to a breakdown 
of the FDT and to long-ranged correlations is a fluid
subject to a constant temperature gradient \cite{Kirkpatrick_Cohen_Dorfman_1982c,
Dorfman_Kirkpatrick_Sengers_1994, Ortiz_Sengers_2007}. Such a system is in a NESS,
so time translational invariance holds exactly. Furthermore, the leading effect of the temperature
gradient turns out to be the coupling between the temperature and the shear velocity in the
Navier-Stokes equations; all other occurrences of the spatially varying temperature can
be replaced by the average temperature. In this leading approximation the system shows
spatial translational invariance as well. 

In the current context, this problem has been discussed in Ref.~\onlinecite{Kirkpatrick_Belitz_2024}.
It is an example of Mechanism II: The externally imposed temperature gradient promotes a term
that is present, but nonlinear, in equilibrium to a linear term in the equations linearized about the
NESS. The relevant observables are (one component of) the shear velocity $u_{\perp} \equiv \delta A_1$ and
the temperature fluctuations $\delta T \equiv \delta A_2$. The inverse propagator is
\be
G^{-1}({\bm k},\omega) = \begin{pmatrix} -i\omega + \nu k^2 & 0 \\
                                                                         \alpha             & -i\omega + \DT k^2
                                                                         \end{pmatrix}
\label{eq:4.29}
\ee
with $\nu$ the kinematic viscosity and $\DT$ the thermal diffusion coefficient. $\alpha$ is a
coupling constant that is proportional to the externally fixed temperature gradient. It is real, even under
parity, and dimensionally a frequency. The fluctuating-force correlations are the same as in 
equilibrium \cite{force_correlations_footnote}, viz.
\be
\Delta({\bm k},\omega) = \begin{pmatrix} 2T\nu{\bm k}^2/\rho & 0 \\
                                                                          0                     & 2T^2 \DT {\bm k}^2/c_p
                                        \end{pmatrix}
\label{eq:4.30}
\ee
with $c_p$ the specific heat at constant pressure.   

For the van Hove functions Eq.~(\ref{eq:3.10}) yields
\bse
\label{eqs:4.31}
\bea
S_{11}({\bm k},\omega) &=& \frac{2T}{\rho}\,\frac{\nu k^2}{\omega^2 + (\nu k^2)^2}\ ,\qquad
\label{eq:4.31a}\\
S_{12}({\bm k},\omega) &=& -\alpha\, \frac{2T}{\rho}\,\frac{\nu k^2}{\omega^2 + (\nu k^2)^2}\,\frac{1}{i\omega + \DT k^2}\ ,\qquad
\label{eq:4.31b}\\
S_{21}({\bm k},\omega) &=& -\alpha\,\frac{2T}{\rho}\,\frac{\nu k^2}{\omega^2 + (\nu k^2)^2}\,\frac{1}{-i\omega + \DT k^2}\ ,\qquad\quad
\label{eq:4.31c}\\
 S_{22}({\bm k},\omega) &=& \frac{2T^2}{c_p}\,\frac{ \DT k^2}{\omega^2 + (\DT k^2)^2}
 \nonumber\\
 && \hskip -20pt + \alpha^2\, \frac{2T}{\rho}\,\frac{1}{\omega^2 + (\DT k^2)^2}\,\frac{\nu k^2}{\omega^2 + (\nu k^2)^2}\ . \qquad
\label{eq:4.31d}\\ 
\nonumber
\eea
\ese
For the static correlation functions, this yields
\bse
\label{eqs:4.32}
\bea
S_{11}({\bm k}) &=& T/\rho\ ,
\label{eq:4.32a}\\
S_{12}({\bm k}) &=& S_{21}({\bm k}) =  \frac{-\alpha T/\rho}{(\nu + \DT)k^2}\ ,
\label{eq:4.32b}\\
S_{22}({\bm k}) &=& \frac{T^2}{c_p} + \frac{\alpha^2 T/\rho}{\DT(\nu + \DT) k^4}\ .
\label{eq:4.32c}
\eea
\ese
Note that these long-ranged correlations result from the coupling between the two diffusive modes in the NESS, {\em not} 
from the random-force correlations. This is in contrast to the case of a randomly stirred fluid, see the remarks after
Eq.~(\ref{eq:4.22}). In an approximation that suffices for capturing the leading terms of the temperature gradient,
Eq.~(\ref{eq:2.6b}) is still valid \cite{Kirkpatrick_Belitz_2024}, but the FDT is not
as can be seen by considering the static response functions. They are given by \cite{static_susceptibility_footnote_2}
\be
X^0({\bm k}) = \begin{pmatrix} \frac{1}{\rho} & \frac{-\alpha}{\rho}\,\frac{1}{\nu k^2} \\
                                                        0           & \frac{T}{c_p} + \frac{\alpha^2}{\rho}\,\frac{1}{\nu \DT k^4}
                                                        \end{pmatrix}\ .
\label{eq:4.33}
\ee
Comparing Eqs.~(\ref{eq:4.33}) and (\ref{eqs:4.32}) we see that the FDT in the form of Eq.~(\ref{eq:2.7a}) is
violated. Note that the non-equilibrium contributions to $X^0$ are strongly singular functions of the wave number
and depend on transport coefficients rather than thermodynamic derivatives 
(see also Ref.~\onlinecite{static_susceptibility_footnote}).
This singular behavior as $k\to 0$, which is also present in the equal-time correlation functions,
Eqs.~(\ref{eqs:4.32}), is the origin of the generalized rigidity and anomalously fast
propagation of perturbations discussed in Refs.~\onlinecite{Kirkpatrick_Belitz_Dorfman_2021}
and \onlinecite{Kirkpatrick_Belitz_2024}.

The response formulas, Eqs.~(\ref{eqs:3.28}), can be constructed from the solutions of the initial-condition problem
\bse
\label{eqs:4.34}
\bea
u_{\perp}^+({\bm k},\omega) &=& \frac{1}{-i\omega + \nu k^2}\,u_{\perp}^0({\bm k}) \ ,
\label{eq:4.34a}\\
\delta T^+({\bm k},\omega) &=& \frac{1}{-i\omega + \DT k^2}\,\delta T^0({\bm k}) 
\nonumber\\
&& - \frac{\alpha}{-i\omega + \nu k^2}\,\frac{1}{-i\omega + \DT k^2}\,u_{\perp}^0({\bm k})\ \qquad\quad
\label{eq:4.34b}
\eea
\ese
that are obtained by using Eq.~(\ref{eq:4.29}) in (\ref{eq:3.20b}). Note that $u_{\perp}^+$ does not depend on the
initial temperature perturbation and is given by the same expression as in equilibrium. For the matrix elements 
of ${\cal A}^{++}$ we obtain
\begin{widetext}
\bse
\label{eqs:4.35}
\bea
{\cal A}_{11}^{++}({\bm k},\omega) &=& \frac{1}{\omega^2 + (\nu k^2)^2}\,\vert u_{\perp}^0({\bm k})\vert^2 \ ,
\label{eq:4.35a}\\
{\cal A}_{12}^{++}({\bm k},\omega) &=&  \frac{1}{-i\omega + \nu k^2}\,\frac{1}{i\omega + \DT k^2}\,u_{\perp}^0({\bm k})\,\delta T^0(-{\bm k})
     - \frac{\alpha}{\omega^2 + (\nu k^2)^2}\,\frac{1}{i\omega + \DT k^2}\,\vert u_{\perp}^0({\bm k})\vert^2\ ,\qquad
\label{eq:4.35b}\\       
{\cal A}_{21}^{++}({\bm k},\omega) &=& \frac{1}{-i\omega + \DT k^2}\,\frac{1}{i\omega + \nu k^2}\,\delta T^0({\bm k})\,u_{\perp}^0(-{\bm k})
     - \frac{\alpha}{-i\omega + \DT k^2}\,\frac{1}{\omega^2 + (\nu k^2)^2}\,\vert u_{\perp}^0({\bm k})\vert^2\ ,\qquad
 \label{eq:4.35c}\\
{\cal A}_{22}^{++}({\bm k},\omega) &=&  \frac{1}{\omega^2 + (\DT k^2)^2}\,\vert\delta T^0({\bm k})\vert^2
       -\frac{\alpha}{i\omega + \nu k^2}\,\frac{1}{\omega^2 + (\DT k^2)^2}\,\delta T^0({\bm k})\,u_{\perp}^0(-{\bm k})   
          \ , \qquad
\nonumber\\  
  &&       -\frac{\alpha}{-i\omega + \nu k^2}\,\frac{1}{\omega^2 + (\DT k^2)^2}\,u_{\perp}^0({\bm k})\,\delta T^0(-{\bm k})
             + \frac{\alpha^2}{\omega^2 + (\nu k^2)^2}\,\frac{1}{\omega^2 + (\DT k^2)^2}\,\vert u_{\perp}^0({\bm k})\vert^2
\ . \qquad
       \label{eq:4.35d}
\eea
\ese
\end{widetext}
Eqs.~(\ref{eqs:4.35}) and (\ref{eqs:4.31}) again demonstrate the isomorphism between ${\cal A}^{++}$
and $S$. In particular, for a pure velocity perturbation, $\delta T^0 = 0$, the non-equilibrium contributions
to the ${\cal A}^{++}_{ij}$ are the same as those to the $S_{ij}$, except for the factors of $k^2$ in the
numerators of the latter, which are a result of the ${\bm k}$-dependence of $\Delta$, Eq.~(\ref{eq:4.30}).

We now consider the response to an external field, which is equivalent to the initial-value problem.
For this, we need the static response functions, Eq.~(\ref{eq:4.33}). 
From Eqs.~(\ref{eqs:4.34}), (\ref{eq:3.15c}), and (\ref{eq:4.33}) we obtain
\bse
\label{eqs:4.36}
\bea
u_{\perp}^+({\bm k},\omega) &=& \frac{1}{\rho}\, \frac{1}{-i\omega + \nu k^2} \left(h_{\perp}({\bm k})
     -  \frac{\alpha}{\nu k^2}\,h_T({\bm k})\right)\ ,
     \nonumber\\
\label{eq:4.36a}\\
\delta T^+({\bm k},\omega) &=& \frac{1}{-i\omega + \DT k^2}
\nonumber\\
&& \hskip -40pt  \times \Biggl\{  \left[\frac{T}{c_p} + \frac{\alpha^2}{\rho}\,\frac{1}{\nu k^2}\left(\frac{1}{\DT k^2} + \frac{1}{-i\omega + \nu k^2}\right)\right]\,  h_T({\bm k})
\nonumber\\
&& \hskip -20pt - \frac{\alpha}{\rho}\,\frac{1}{-i\omega + \nu k^2}\,\frac{1}{-i\omega + \DT k^2}\,h_{\perp}({\bm k}) \Biggr\}\ .
\label{eq:4.36b}
\eea
\ese
Here $h_{\perp}$ and $h_T$ are the fields conjugate to the shear velocity and the temperature, respectively.
We note that $u_{\perp}^+$ depends on $h_T$ as well as on $h_{\perp}$, whereas it does {\em not} depend on
$\delta T^0$, see Eq.~(\ref{eq:4.34a}). This is an example of the effects of a non-diagonal $X^0$ that were mentioned at the
end of Sec.~\ref{subsec:III.B}.

Forming bilinear products of $u_{\perp}^+$ and $\delta T^+$ as expressed in Eqs.~(\ref{eqs:4.36}) yields 
the matrix elements ${\cal A}_{ij}$ in terms of the fields $h_{\perp}$ and $h_T$. For the special case $h_T = 0$ 
we recover expressions that are equivalent to those given in Ref.~\onlinecite{Kirkpatrick_Belitz_2024}.

\section{Summary, and Discussion}
\label{sec:V}

In summary, we have presented a relation between correlation functions and response or relaxation
functions in many-body systems that remains valid in non-equilibrium states where the usual
fluctuation-dissipation theorem (FDT) breaks down. It takes the form of an isomorphism between the
two-point correlation functions and bilinear products of linear response formulas. The technical
underpinning is the fact that fluctuating forces, which determine the correlations, and external
fields, which determine the response, both enter as linear inhomogeneities in a generalized
Langevin equation. The practical significance lies in the fact that fluctuation formulas and
response formulas contain the same information about correlations, but the latter are amenable
to experimental probes even in situations where the direct measurement of correlation functions
is very difficult. 

In what follows we present various discussion points, some of which elaborate on remarks
that were already made in the main text.

\begin{enumerate}[leftmargin=*]
\item We have used a generalized Langevin equation, Eqs.~(\ref{eqs:3.1}), that is
obtained by linearizing about an arbitrary state that can be an equilibrium or non-equilibrium
state, and may or may not be time dependent. Within this formalism the crucial isomorphism,
expressed in Sec.~\ref{subsec:III.C}, manifests itself in a very simple way. It is an obvious
result of the fluctuating forces and the conjugate fields both being linear inhomogeneities in
the Langevin equation, and it is arguably more plausible than the usual fluctuation-dissipation 
theorem. This simplicity is somewhat deceptive, as the result is much less obvious if one uses 
different techniques. In this context it is useful to remember that in non-equilibrium states the 
dynamics play a much more crucial role than in equilibrium ones. For instance, in non-equilibrium 
states the equal-time correlation functions in general depend on the dynamics (see, for instance,
Eqs.~(\ref{eqs:4.32})), whereas in 
equilibrium they are thermodynamic quantities that are entirely determined by the partition 
function. The use of dynamic equations is therefore very natural in this context.

A related point is that the isomorphism reflects Onsager's regression hypothesis \cite{Onsager_1931b},
as was mentioned at the end of Sec.~\ref{sec:III}. Actually, the very fact that one can
use the dynamic equation (\ref{eq:3.1a}) to discuss both fluctuations and response or relaxation
relies on Onsager's hypothesis. In the language of the current paper, it states that the relaxation 
of the dynamic variables from an initial value is governed by the averaged version of the same 
dynamic equation that governs the dynamics of the spontaneous fluctuations.
As a practical consequence, the correlation functions even in non-equilibrium systems can
effectively be determined by measuring response or relaxation functions, and the size of the
latter is given by macroscopic driving forces that are experimentally controllable even if the
fluctuations are small. This point was made in the context of the example of a fluid in a
NESS, Sec.~\ref{subsec:IV.C}, in Refs.~\onlinecite{Kirkpatrick_Belitz_2023b, Kirkpatrick_Belitz_2024}.                  
\item What constitutes a violation of the FDT is to some extent a matter of choice. The
most common notion, which we have adopted, is that Eqs.~(\ref{eqs:2.16})
must hold as {\em equalities}, as opposed to just proportionality relations, in order for
the FDT to hold. We have discussed two mechanisms for violating the FDT, see 
Sec.~\ref{sec:IV}. Mechanism I is very simple: it consists of stipulating a fluctuating-force 
correlation that does not lead to the equilibrium equal-time correlation functions. While
technically trivial, this reflects the fact that the equal-time correlation functions are in
general not known in non-equilibrium systems and depend on the dynamics. Since the 
response functions do not depend on the fluctuating-force correlations this obviously 
leads to a violation of the FDT. This simple but fundamental point implies that an 
equilibrium equal-time correlation function that is different from the equilibrium one implies a 
violation of the FDT. Mechanism I does, however, leave intact the structure of the
equilibrium theory. For Mechanism I (a) in particular, where the fluctuating-force correlation
has the same wave-number dependence as in equilibrium, the FDT is violated only in the 
sense that Eq.~(\ref{eq:2.16b}) turns into a proportionality relation with a constant 
proportionality factor. One might call this a weak violation of the FDT. Examples we have 
discussed are active Brownian motion, Sec.~\ref{subsubsec:IV.B.1}, and the randomly 
stirred fluid, Sec.~\ref{subsubsec:IV.B.2}, with a $\Delta({\bm k}) \propto k^2$
that is not given by Eq.~(\ref{eq:4.2}).  In these cases the fluctuation-response relation 
from Sec.~\ref{subsec:III.C} is no stronger a statement than the violated FDT, since it takes the form of an
isomorphism rather than an identity. However, we emphasize again that the isomorphism
{\em always} holds, irrespective of whether, or how badly, the FDT is
violated. 

For Mechanism I (b) the above structural remarks are still true, but the wave-number
dependence of the fluctuating-force correlations can lead to long-range correlations that are
not present in equilibrium. These long-range correlations can be spatial in nature, as in the
example of a stirred fluid with $\Delta({\bm k}) = \text{const.}$, Eq.~(\ref{eq:4.22}), or temporal,
as in the example of cell motion in Sec.~\ref{subsubsec:IV.B.3}. In these examples the
equal-time correlation function scales differently with the wave number than the static
response function. This is a strong violation of the FDT. 

Mechanism II consists of couplings between the dynamic variables that are
different from the equilibrium theory, which leads to structural changes. Of our
examples, the fluid in a NESS, Sec.~\ref{subsec:IV.C}, falls into this category. In this
case the equal-time correlation functions are proportional to the static response
functions, see Eqs.~(\ref{eqs:4.32}, \ref{eq:4.33}), although the non-equilibrium
contributions to both are singular functions of the wave number and the proportionality
factors depend on kinetic coefficients. Also, one of the response functions vanishes
whereas the corresponding equal-time correlation functions is nonzero. This is in
between the weak and strong violations of the FDT as defined above. 

It is possible to have both mechanisms present, although this is not the
case in any of the examples we have considered. This is an interesting problem to consider. 
\item In the context of Mechanism II, which relies on terms that are absent in equilibrium
hydrodynamics, it is worthwhile mentioning that terms that {\em are} present in ordinary
hydrodynamics, but are anomalously large due to a non-equilibrium situation, can lead
to qualitative effects. An example is the model of active particles suspended in a 
momentum-conserving fluid studied in Ref.~\onlinecite{Kirkpatrick_Bhattarcherjee_2019},
where an anomalously large Burnett coefficient leads to a hydrodynamic instability.
\item In equilibrium statistical mechanics, the static response functions $X^0$ are naturally
considered as input into the linear-response problem. In this case, they are purely
thermodynamic quantities and are given by commutator correlation functions $\chi^0$. In
non-equilibrium systems none of the preceding statements are true. In the context of the
dynamic equations (\ref{eqs:3.1}), it is the $b_{ij}$ that describe the coupling of the
variables to the external fields, and it is natural to consider them fundamental
ingredients of the theory on the same level as the couplings and relaxation rates
$\lambda_{ij}$. The static response functions $X^0$ then are derived quantities
that are related to the fundamental couplings $b_{ij}$ via Eq.~(\ref{eq:3.21b}).
Even though the $X^0$ are not frequency dependent, in a non-equilibrium system
they in general depend on the dynamics via kinetic coefficients, as we have seen
explicitly in Sec.~\ref{subsec:IV.C}, see Eq.~(\ref{eq:4.33}). We also note that the
frequency dependence of $b$ is a consequence of the frequency dependence of
$\lambda$, i.e., of memory effects. If one neglects memory effects, both $\lambda$
and $b$ are delta functions in time space and hence frequency independent.
\item In all of our examples the Langevin-force correlations were short ranged. This raises
the question as to whether long-ranged correlations ever arise from long-ranged stochastic
forces. This is to some degree a matter of how the theory is formulated. Consider the
example of a fluid in a NESS, Sec.~\ref{subsec:IV.C}. As was shown in 
Ref.~\onlinecite{Kirkpatrick_Belitz_Dorfman_2021}, it is possible to formulate a theory for
the temperature fluctuations only, while the velocity fluctuations are hidden by integrating
them out in a Martin-Siggia-Rose formalism. The resulting effective Langevin equation
for the temperature fluctuations has a long-ranged stochastic force. This is analogous to
the well known phenomenon in field theory, where integrating out soft, or massless, modes
leads to long-ranged effective interactions. Conversely, a long-ranged Langevin force can
be made short-ranged by introducing suitable auxiliary variables that make explicit the
`hidden sector' of the theory. 
\item For our discussion of response functions we have assumed time translational
invariance. This is an exact property of steady states, but for general non-equilibrium
states it is true at best in some approximate sense. An exploration of the conditions
under which such an approximation is valid, as well as a possible generalization of the 
fluctuation-response relation to situations where time-translational invariance does not
hold, are interesting problems. 

With this proviso in mind we stress again the generality of the fluctuation-response
relation derived in Sec.~\ref{subsec:III.C}. For instance, the large fluctuations present 
in the NESS discussed in Sec.~\ref{subsec:IV.C} can give rise to instabilities such
as Rayleigh-B{\'e}nard convection \cite{Ortiz_Sengers_2007}. This leads to a new NESS that may,
as a function of suitable parameters, become unstable itself and ultimately lead to a time-dependent
state. The fluctuation-response relation will remain valid in such a scenario if one linearizes 
about the appropriate state. Similar remarks hold for instabilities in active matter systems,
see Ref.~\onlinecite{Kirkpatrick_Bhattarcherjee_2019}, and also Ref.~\onlinecite{Marchetti_et_al_2013} 
and references therein.
\end{enumerate}

\hrule
\bigskip

\appendix
\section{Causal functions}
\label{app:A}

Let $g(t)$ be a time-dependent function. It is useful to separate $g$ into parts for positive and
negative times, respectively,
\bse
\label{eqs:A.1}
\be
g(t) = g^-(t) + g^+(t)\ ,
\label{eq:A.1a}
\ee
with
\be
g^{\pm}(t) = \Theta(\pm t)\,g(t)\ .
\label{eq:A.1b}
\ee
\ese
We define a temporal Fourier transform
\bse
\label{eqs:A.2}
\be
g(\omega) = \int_{-\infty}^{\infty} dt\,e^{i\omega t}\,g(t)\ ,
\label{eq:A.2a}
\ee
and a Laplace transform
\be
g(z) = {\pm\int_{-\infty}^{\infty} dt\,e^{izt}\,g^{\pm}(t)}
 \qquad (\pm\ \text{for}\ \text{Im}(z)  \genfrac{}{}{0pt}{2}{>}{<} 0)\ ,
\label{eq:A.2b}
\ee
with $z$ the complex frequency. (We denote the underlying real-time function, its Fourier transform,
and its Laplace transform, by the same symbol and distinguish between them by their arguments.)
With these definitions,
\be
g^{\pm}(\omega) = \pm g(z=\omega\pm i\epsilon)\ ,
\label{eq:A.2c}
\ee
with $\epsilon>0$ infinitesimal. The Fourier transform is related to the discontinuity 
across the cut on the real frequency axis, which defines the spectral density $g''$
\bea
g''(\omega) &=& \frac{1}{2i}\,[g(z=\omega+i\epsilon) - g(z=\omega-i\epsilon)]
\nonumber\\
                  &=& g(\omega)/2i\ .
\label{eq:A.2d}
\eea
The spectral density in turn is related, via the Sokhotski-Plemlj formula, to the Laplace transform by the Hilbert-Stieltjes 
relation \cite{Muskhelishvili_2013}
\be
g(z) = \int_{-\infty}^{\infty} \frac{d\omega}{\pi}\,\frac{g''(\omega)}{\omega - z}\ .
\label{eq:A.2e}
\ee
\ese


\begin{thebibliography}{60}
\bibitem[{tim()}]{time_translation_invariance_footnote}
\bibinfo{note}{Time translational invariance always holds in a steady, or
  stationary, state, which may or may not be an equilibrium state. More
  generally, it may hold approximately if the time dependence of the state is
  slow in an appropriate sense. This point requires additional investigations,
  see Sec.~\ref{sec:IV}.}
\bibitem[{cau()}]{causality_footnote}
\bibinfo{note}{Causality restricts the response function to positive times. As
  a result, $X_{ij}$ for negative times and its Laplace transform in the lower
  half plane have no physical meaning.}
\bibitem[{lin()}]{linearizing_footnote}
\bibinfo{note}{By definition, linearizing to describe the linear response to an
  external field, or to imposed initial conditions, is justified. For the
  thermal correlation functions it is known that nonlinear terms in the
  fluctuations simply renormalize quantities such as the transport coefficients
  in the macroscopic equations. In general, this is a small effect. However,
  this may not be true near critical or instability points, or if the noise
  correlation is anomalously large, see Sec.~\ref{subsubsec:IV.B.3}.}
\bibitem[{pro()}]{propagator_footnote}
\bibinfo{note}{$G$ and $G^{\text{T}}$ are in general operators, and
  Eq.~(\ref{eq:3.6}) needs to be interpreted accordingly.}
\bibitem[{b_f()}]{b_footnote}
\bibinfo{note}{For a given propagator, knowledge of the coupling constants $b$
  is equivalent to knowledge of the static response functions $X^0$ (which in
  general depend on the dynamics). Within the framework of the dynamic
  equations the $b$ should be considered fundamental quantities, and the $X^0$
  should be considered derived ones. We will further discuss this in
  Sec.~\ref{sec:V}.}
\bibitem[{act()}]{active_matter_footnote}
\bibinfo{note}{The term `active matter' refers to systems governed by
  statistical mechanics that are not in equilibrium due to intrinsic energy
  sources. This is a vast field, and we briefly discuss only a few examples
  that illustrates the general point made in this paper.}
\bibitem[{sta({\natexlab{a}})}]{static_susceptibility_footnote}
\bibinfo{note}{Note that the static susceptibility must be determined
  independently, it cannot be obtained from the sum rule (\ref{eq:2.7b}). Also
  note that in equilibrium the static susceptibility is a thermodynamic
  quantity, whereas in non-equilibrium systems it requires knowledge of the
  dynamics.}
\bibitem[{upe()}]{uperp_footnote}
\bibinfo{note}{In this simple model the velocity can rotate freely, so the
  equation of motion for the velocity components perpendicular to $\langle{\bm
  v}\rangle_{h=0}$ is not physically meaningful.}
\bibitem[{sta({\natexlab{b}})}]{static_susceptibility_footnote_2}
\bibinfo{note}{This result can be obtained from the time-dependent
  Ginzburg-Landau theory known as Model H in
  Ref.~\onlinecite{Hohenberg_Halperin_1977}. It also has been derived from
  kinetic theory in Ref.~\onlinecite{Kirkpatrick_Belitz_2023a}. Both of these
  derivations yield expressions for the quantity $b_{ij}$,
  Eq.~(\ref{eq:3.21b}). $X^0$ is obtained by using the propagator in that
  equation and solving for $X^0$.}
\end{thebibliography}

\end{document}